\documentstyle[12pt]{article}
\def\singlespace {\smallskipamount=3.75pt plus1pt minus1pt
                  \medskipamount=7.5pt plus2pt minus2pt
                  \bigskipamount=15pt plus4pt minus4pt
                  \normalbaselineskip=15pt plus0pt minus0pt
                  \normallineskip=1pt
                  \normallineskiplimit=0pt
                  \jot=3.75pt
                  {\def\smallskip {\vskip\smallskipamount}}
                  {\def\medskip   {\vskip\medskipamount}}
                  {\def\bigskip   {\vskip\bigskipamount}}
                  {\setbox\strutbox=\hbox{\vrule
                    height10.5pt depth4.5pt width 0pt}}
                  \parskip 7.5pt
                  \normalbaselines}
\def\middlespace {\smallskipamount=5.625pt plus1.5pt minus1.5pt
                  \medskipamount=11.25pt plus3pt minus3pt
                  \bigskipamount=22.5pt plus6pt minus6pt
                  \normalbaselineskip=22.5pt plus0pt minus0pt
                  \normallineskip=1pt
                  \normallineskiplimit=0pt
                  \jot=5.625pt
                  {\def\smallskip {\vskip\smallskipamount}}
                  {\def\medskip   {\vskip\medskipamount}}
                  {\def\bigskip   {\vskip\bigskipamount}}
                  {\setbox\strutbox=\hbox{\vrule
                    height15.75pt depth6.75pt width 0pt}}
                  \parskip 11.25pt
                  \normalbaselines}
\def\doublespace {\smallskipamount=7.5pt plus2pt minus2pt
                  \medskipamount=15pt plus4pt minus4pt
                  \bigskipamount=30pt plus8pt minus8pt
                  \normalbaselineskip=30pt plus0pt minus0pt
                  \normallineskip=2pt
                  \normallineskiplimit=0pt
                  \jot=7.5pt
                  {\def\smallskip {\vskip\smallskipamount}}
                  {\def\medskip   {\vskip\medskipamount}}
                  {\def\bigskip   {\vskip\bigskipamount}}
                  {\setbox\strutbox=\hbox{\vrule
                    height21.0pt depth9.0pt width 0pt}}
                  \parskip 15.0pt
                  \normalbaselines}
\begin{document}
\title{Three lepton decay mode of the proton}
\author{Patrick J. O'Donnell \\
Physics Department \\
University of Toronto \\
Toronto, Ontario M5S 1A7, Canada\\
\\ and \\ \\
Utpal Sarkar \\
Theory Group \\
Physical Research Laboratory \\
Ahmedabad - 380 009, India}

\date{\mbox{}}

\maketitle

\begin{abstract}
\middlespace

We consider the three lepton decay modes of the proton within the proton decay
interpretation of the atmospheric neutrino anomaly.  We construct higher
dimensional operators in the framework of the standard model.  The operators
which allow the interesting decay modes are of dimension 10 involving $SU(2)_L$
non-singlet higgs.  We show how these operators can be comparable to the
dimension 9 operators.  We then present a simple model which can give rise to
the desired proton decay modes of the right order of magnitude.

\end{abstract}
\newpage
\middlespace

Muon  neutrinos  produced by cosmic  rays in the  atmosphere  are
expected to be almost twice as many as electron  neutrinos (where
neutrinos  are not  distinguished  from  antineutrinos).  But the
results   \cite{atm1,atm2}  from  the  two  large  water-Cerenkov
detectors  are  $R_{obs}/R_{MC}=0.60  \pm 0.07 \pm 0.05$ from the
Kamiokande experiment  \cite{atm1} and  $R_{obs}/R_{MC}=0.54  \pm
0.05 \pm 0.12$ from IMB  \cite{atm2},  {\it  i.e.,} the  observed
ratio $R =  N(\nu_\mu)/N(\nu_e)$  is  almost  half  the  expected
ratio.  The experiments look for  ``contained''  events which are
caused by  neutrinos  of energy  below 2 GeV.  Although  the more
popular  explanation  of this  anomaly  is  neutrino  oscillation
\cite{atmrev},  there is another  explanation  in terms of proton
decay  \cite{mann}.  It has been  proposed  that if proton decays
into a positron  and two  neutrinos  with a lifetime of $4 \times
10^{31}$ years (this value is consistent  with the present proton
decay  limit for this decay mode  \cite{taup}),  then the  excess
``contained''  electron  events  can  actually  be  proton  decay
events.  Since the energy of these  electrons peak around 350 MeV
with a distribution ranging up to 1 GeV, the decay mode has to be
$P \to e^+ \nu  \nu$.  We are not  interested  in  neutron  decay
since  neutron  decay  events  cannot  explain  the   atmospheric
neutrino  anomaly; neutron decay events have to have at least two
charged  leptons  and hence  give two leg  ``contained''  events,
which  have  not been  observed.  The  possibility  of the  three
lepton  decay  mode for the  proton  has been  discussed  earlier
\cite{pss,pati},   but  it  is  difficult  to  incorporate   this
particular decay mode in those theories \cite{rudaz}.

For this  mechanism  to work in any  theoretical  model, the main
problem is to have this decay mode with only light neutrinos.  In most
theories left handed  neutrinos are light and right handed
neutrinos are heavy.  The decay modes are thus  restricted  to
\begin{equation}
P \to e^+ \nu_L \nu_L \;\;\;\; {\rm or} \;\;\;\; P \to e^+
{\nu_L}^c  {\nu_L} \;\;\;\; {\rm or} \;\;\;\; P \to e^+
{\nu_L}^c  {\nu_L}^c.  \label{dkmod}
\end{equation}
These     processes     require     six     fermion     operators
\cite{pss,pati,rudaz}  and  hence  are  usually  more  suppressed
\cite{gutrev}  than decay modes of the type $P \to e^+ \pi^0$
(for which  $\tau_P > 5 \times  10^{32}$  years  \cite{tauprot}),
which  are  allowed  by four  fermion  operators.  Thus  the next
problem is to make  (\ref{dkmod})  more  dominate  the
decay modes.

In this  article  we study this decay  mode in  detail.  We first
make   an    operator    analysis    \cite{rudaz,weinberg}    for
(\ref{dkmod}).  We write down the  effective  operators of higher
dimension $n$ allowed by the standard model, which are suppressed
by  $M^{(n-4)}$,  where $M$ is the mass scale in the  theory, and
which depends on the details of the model.  We consider the higgs
scalars which break the standard  model,  namely, a higgs doublet
and a higgs  triplet.  Although a triplet is not  present  in the
minimal   standard  model  it  naturally   exists  in  left-right
symmetric theories \cite{lrm}.

We construct  operators  involving  only the known  fermions  and
which are invariant under the standard model gauge group $SU(3)_c
\otimes  SU(2)_L  \otimes  U(1)_Y$.  In  ref  \cite{rudaz}  these
operators (not including higgs scalars) in the form $QQQLLL$ were
discussed.  We review these  operators in some detail since there
are  subtleties   that  are   important.  Consider  the  operator
$[\overline{{\psi_{iL}}^c}  \psi_{jL}]  [\overline{{\psi_{kL}}^c}
\psi_{lL}]$, antisymmetric in $[ij]$, which vanishes if the $i$th
and the $j$th particles  belong to the same  generation.  However
the      operator      $[\overline{{\psi_{iL}}^c}      \psi_{lL}]
[\overline{{\psi_{kL}}^c}  \psi_{jL}]$  may be  antisymmetric  in
$[ij]$ and non-vanishing for one generation, even though both the
operators  look like $\psi_L  \psi_L  \psi_L \psi_L $ with two of
them  antisymmetric.  In  fact, if $A$ and $B$ are the  same  two
fields which enter  antisymmetrically  in an operator ${\cal O} =
[\overline{A^c} B][\overline{C^c} D]$, then the operator vanishes
only when the other fields $C$ and $D$ are of different  helicity
from  $A$  and  $B$.  Otherwise  one  can  always  write  another
operator ${\cal O}' = [\overline{A^c} D][\overline{C^c} B]$ which
can be Fierz  transformed  to a combination  of ${\cal O}$ and an
operator of the form $[\bar{A^c}  \sigma_{\mu  \nu}  B][\bar{C^c}
\sigma^{\mu \nu} D]$, which is non-vanishing  for one generation.
This  is  because  $$  \epsilon_{i  j}   (\overline{\psi_{iaL}^c}
\psi_{jbL})   =   -   \epsilon_{i   j}   (\overline{\psi_{ibL}^c}
\psi_{jaL})     \;\;{\rm    and    }    \;\;    \epsilon_{i    j}
(\overline{\psi_{iaL}^c}    \sigma_{\mu   \nu}    \psi_{jbL})   =
\epsilon_{i   j}   (\overline{\psi_{ibL}^c}    \sigma^{\mu   \nu}
\psi_{jaL}) .$$

We now write the detailed form of the  operators  and discuss the
possibility  of the  decay  mode  (\ref{dkmod}).  We  work in the
context of  conventional  models.  This means that the  neutrinos
are  left-handed.  In the standard model there is no right handed
neutrino  and in  left-right  symmetric  models the right  handed
neutrinos   are  too  heavy.  The  other  fields  can  be  either
left-handed or right handed.

There are only four dimension nine operators  involving six
fermions,    which   could  allow    (\ref{dkmod}).   These    are
\begin{eqnarray}
{{\cal  O}^1} & = &  (\overline{{q_{i  \alpha a
L}}^c} q_{j \beta b L})  (\overline{{d_{\gamma  c R}}^c} e_{d R})
(\overline{l_{k  e L}}  {l_{l  f  L}}^c)  \epsilon_{\alpha  \beta
\gamma} (\tau^I \epsilon)_{i j} (\tau^I \epsilon)_{k l}
	\nonumber \\
{{\cal O}^2} & = & (\overline{{q_{i \alpha a L}}^c} q_{j \beta
b L})(\overline{{d_{\gamma c R}}^c} {l_{k d L}}^c)(\overline{l_{l
e  L}}  e_{f   R})   \epsilon_{\alpha   \beta   \gamma}   (\tau^I
\epsilon)_{i j} (\tau^I \epsilon)_{k l}
	\nonumber \\
{{\cal O}^3} & = & (\overline{{u_{\alpha a R}}^c}  u_{\beta b R})
(\overline{{q_{i  \gamma c L}}^c}  l_{j d L})  (\overline{l_{k  e
L}}^c  {l_{l  f  L}})  \epsilon_{\alpha   \beta  \gamma}  (\tau^I
\epsilon)_{i j} (\tau^I \epsilon)_{k l}
	\nonumber \\
{{\cal O}^4} &=&(\overline{{u_{\alpha a R}}^c} u_{\beta b R})
(\overline{{q_{i  \gamma c L}}^c}  l_{j d L})  (\overline{l_{k  e
L}}^c {l_{l f L}}) \epsilon_{\alpha  \beta \gamma}  (\tau^I
\epsilon)_{i j} (\tau^I \epsilon)_{k l} \label{op1}
\end{eqnarray}

\noindent where, $\alpha, \beta, \gamma $ are $SU(3)_c$  indices,
$i, j, k, ...  [I, J,  ...]$  are  $SU(2)_L$  doublet  [triplet]
indices  and $a, b, c, ...$  are  generation  indices.  Since  by
Fierz  transformation  the gauge boson mediated operators of type
$(\overline{\psi_L}    \gamma_\mu   \psi_R)    (\overline{\psi_L}
\gamma^\mu \psi_R)$ are contained in $(\overline{\psi_L}  \psi_L)
(\overline{\psi_R}  \psi_R)$ we have not written them separately.

In ${{\cal O}^1}$ and ${{\cal O}^2}$ the left handed quarks $q_L$
have to be up quarks.  In all the four  operators  the  up-quarks
are  antisymmetric  in the $SU(3)_c$ index while symmetric in all
other indices so they have to be  antisymmetric in the generation
indices.  As a result all of these  operators  can only give rise
to proton  decay  into  charmed  mesons,  which is  kinematically
forbidden.

For an  antisymmetric  combination of left-handed  [right-handed]
fields to be  non-vanishing,  two more  left-handed[right-handed]
fields are  necessary as discussed  earlier.  In the present case
the neutrino fields have opposite helicity to the up quarks since
$({\psi_L})^c =
({\psi^c})_R$, $\overline{(\psi_R)} = (\overline{\psi})_L$
and $\overline{({\psi_L})^c} = \overline{({{\psi^c})_R}} =
(\overline{{{\psi^c}}})_L$ and
hence  $(\overline{{q_L}^c}  {l_L}^c) = 0$.  Thus there cannot be
any term of the form $(\overline{{q_L}^c} {l_L}^c)(\overline{l_L}
q_L)$  except the gauge boson  mediated  ones, which can be Fierz
transformed to the operators we have already  listed.  Unless one
can  make  a  model  which  has a  light  right  handed  neutrino
\cite{rudaz}  it is not  possible  to have the proton  decay mode
(\ref{dkmod}) without higgs scalars.

We now  study  the  higher  dimensional  operators  which  are
possible with those higgs scalars whose vacuum expectation values
({\it  vev}\,s)  give rise to the decay modes  (\ref{dkmod}).  We
consider two types of higgs scalars which are non-singlets  under
the  standard  model.  One type is the  usual  $SU(2)_L$  doublet
field  $\phi$,  which  gives  masses to the  fermions.  The other
possibility  is a $SU(2)_L$  triplet  $\Delta_L$  in  addition to
$\phi$.  This field  $\Delta_L$  is always  present in left-right
symmetric  theories and can give Majorana mass to the left-handed
neutrinos.

The operators without the higgs scalars are of dimension 9 and so
this decay mode is suppressed  by a factor $M^5$ in the amplitude
where $M$ is the mass scale in the theory.  Thus,
\begin{equation}
\tau_P \sim  \:\:\: M^{10}
\end{equation}
and for the experimental value \cite{taup}, $\tau_P \sim 4 \times
10^{31}$  yrs, we find $M \sim 10^6$ GeV.  A  dimension  10
operator  can give rise to these  decay
modes when the higgs bosons $\phi$ or $\Delta_L$  acquires a {\it
vev}, $\eta$, say.   The
amplitude   is   suppressed   by   a   factor   of
${M^6/\eta}$,  {\it i.e.}, the lifetime for the process
becomes,
\begin{equation}
\tau_P  \sim   \:\:\:
\displaystyle  \frac{M^{12}}{{  \eta}^2} .
\end{equation}
For a doublet  field $\eta = \langle  \phi \rangle \sim 250$ GeV,
while for a triplet  field\footnote{The  upper bound on $v_L \sim
O(1)$ GeV comes  from LEP data  \cite{lep90}.}  it can be $\eta =
\langle  \Delta_L  \rangle = v_L \sim O(1)$ GeV, which implies $M
\sim 10^5$, not much  different from the dimension 9 case.  Hence
these dimension 10 operators can be as important as the dimension
9  operators;  we present  below an  explicit  model  where, with
reasonable  choices of  parameters,  such  dimension 10 operators
give rise to the desired proton decay mode (\ref{dkmod}).

We first consider the doublet field $\phi$  transforming  as (2, -1)
under  $SU(2)_L  \times  U(1)_Y$.  Starting  with a  dimension  9
operator  (such as $\psi_L  \psi_L \psi_R \psi_R \psi_L  \psi_L$)
one can contract the $SU(2)_L$ index of a left-handed  field with
that of the $\phi$.  Then one of the remaining $\psi_L$ has to be
replaced by a right-handed  field for $SU(2)_L$  invariance.
The   bilinear   forms  in  any  operator  can  be  of  the  form
$\overline{{\psi_L}^c} \psi_L$,  $\overline{{\psi_R}^c}  \psi_R$,
$\overline{{\psi_L}}  \psi_R$  or  $\overline{{\psi_R}}  \psi_L$.
Now  if  only  one  of  the   left-handed   fields  changes  to  a
right-handed   field  then  this  one,  or  at  least  one  other
left-handed  field,  has  to be  charge  conjugated  to  keep  the
bilinear form from vanishing.  For example,  ${{\psi_L}}  \psi_L$
can be replaced by ${{\psi_R}}  {\psi_L}^c \phi$ or ${{\psi_R}^c}
{\psi_L}  \phi$.  This is because the $\psi_L$ and  ${\psi_L}^c =
{\psi^c}_R$   are   $SU(2)_L$   doublets,   while   $\psi_R$  and
${\psi_R}^c  =  {\psi^c}_L$  are  singlets.  Since only a neutral
component   of  $\phi$  can  acquire  a  {\it  vev},  by charge
conservation  only a neutrino  field can  undergo  such  helicity
flip.  With this criteria we can now write down a set of dimension
10 operators involving the higgs doublet $\phi$ as

\vbox{
\begin{eqnarray}
{\cal O}^1 & = & (\overline{{q_{i
\alpha a L}}^c} q_{j \beta b L})(\overline{{d_{\gamma c R}}^c}
{l_{k d L}}^c)(\overline{{l_{m e L}}^c} l_{l f L}) \phi_n
\epsilon_{\alpha \beta \gamma} (\tau^I \epsilon)_{i j}
(\tau^I \epsilon)_{k l} \epsilon_{m n}
      \nonumber \\
{\cal O}^2 & = & (\overline{{q_{i
\alpha a L}}^c} l_{j b L})(\overline{{d_{\beta c R}}^c}
{l_{m d L}}^c)(\overline{{l_{k e L}}^c} q_{l \gamma f L}) \phi_n
\epsilon_{\alpha \beta \gamma} (\tau^I \epsilon)_{i j}
(\tau^J \epsilon)_{k l} (\tau^K \epsilon)_{m n}
(\tau^K \epsilon)_{I J}
      \nonumber \\
{\cal O}^3 & = & (\overline{{q_{i
\alpha a L}}^c} q_{j \beta b L})(\overline{{q_{k \gamma c L}}^c}
{l_{m d L}})(\overline{l_{l e L}} e_{f R}) \phi_n
\epsilon_{\alpha \beta \gamma} (\tau^I \epsilon)_{i j}
(\tau^I \epsilon)_{k l} \epsilon_{m n}
      \nonumber \\
{\cal O}^4 & = & (\overline{{q_{i
\alpha a L}}^c} l_{m b L})(\overline{{q_{k \beta c L}}^c}
{q_{l \gamma d L}})(\overline{l_{j e L}} e_{f R}) \phi_n
\epsilon_{\alpha \beta \gamma} (\tau^I \epsilon)_{i j}
(\tau^I \epsilon)_{k l} \epsilon_{m n}
      \nonumber \\
{{\cal O}^5} & = &  (\overline{{u_{\alpha a R}}^c} u_{\beta
b R}) (\overline{{d_{\gamma c R}}^c} {l_{i d L}}) (\overline{l_{k
e L}} {l_{j f L}}) \phi_l^\dagger \epsilon_{\alpha \beta \gamma}
(\tau^I \epsilon)_{i j}  (\tau^I \epsilon)_{k l}
      \nonumber \\
{{\cal O}^6} & = &  (\overline{{u_{\alpha a R}}^c} l_{i
b L}) (\overline{{d_{\gamma c R}}^c} u_{\beta
d R}) (\overline{{l_{k e L}}} l_{j f L}) \phi_l^\dagger
\epsilon_{\alpha \beta \gamma} (\tau^I \epsilon)_{i j}
(\tau^I \epsilon)_{k l}
      \nonumber \\
{{\cal O}^7} & = &  (\overline{{u_{\alpha a R}}^c} u_{\beta
b R}) (\overline{{q_{i \gamma c L}}^c} l_{j d L}) (\overline{l_{k
e L}} {e_{f R}}) \phi_l^\dagger  \epsilon_{\alpha \beta \gamma}
(\tau^I \epsilon)_{i j}  (\tau^I \epsilon)_{k l}
      \nonumber \\
{{\cal O}^8} & = &  (\overline{{u_{\alpha a R}}^c} e_{b R})
(\overline{{q_{i \beta c L}}^c} l_{j d L}) (\overline{l_{k
e L}} u_{\gamma f R}) \phi_l^\dagger \epsilon_{\alpha \beta \gamma}
(\tau^I \epsilon)_{i j}  (\tau^I \epsilon)_{k l}
      \label{op2}
\end{eqnarray}}

\noindent From these dimension 10 operators it is clear
that the higgs  $\phi$  allows only $(B - L)$  conserving  proton
decays.  The $SU(2)$ indices have been contracted in only one way
for each of these  operators.  There are  other  operators  which
differ in the way the $SU(2)$  indices  are  contracted.  However
the main  features of these  operators  are  present in all other
operators.  In particular  operators  ${\cal O}^1$, ${\cal O}^3$,
${\cal  O}^5$  and  ${\cal  O}^7$  vanish  for  one   generation,
irrespective of their  $SU(2)_L$  contraction,  {\it i.e.,} these
operators  do not give the decay modes  (\ref{dkmod}).  The other
operators still allow for the decays.  Both the operators  ${\cal
O}^1$ and ${\cal O}^2$ are of the form  $QQd_RLL\bar{L}\phi$  and
similarly   ${\cal  O}^3$  and  ${\cal  O}^4$  are  of  the  form
$QQQe_rL\bar{L}\phi$;  ${\cal  O}^5$ and ${\cal  O}^6$ are of the
form   $u_Ru_Rd_RLL\bar{L}\phi^\dagger$;  and  ${\cal  O}^7$  and
${\cal  O}^8$  are of the form  $u_Ru_RQe_RL\bar{L}\phi^\dagger$.
But  while  ${\cal  O}^n$  (with  odd $n$) can not give the decay
modes  (\ref{dkmod}), the ${\cal O}^{(n+1)}$  operators can allow
decays of (\ref{dkmod}).

In the minimal  left-right  symmetric  model  \cite{pss,pati,lrm}
there exists a natural  choice of the higgs  scalar  masses which
lets us develop a simple model which gives one of the decay modes
of  (\ref{dkmod})  with the right  order of  magnitude.  The {\it
vev}  of  the  right  handed  triplet  higgs  field  $\Delta_R  $
(1,1,3,-2), breaks $G_{LR} \equiv SU(3)_c \otimes SU(2)_L \otimes
SU(2)_R   \otimes   U(1)_{B-L}$  to  the  standard   model.  From
left-right  parity  ($D-$parity)  there  is a  $\Delta_L  $ which
transforms  as  (1,3,1,-2)  under  $G_{LR}$.  The standard  model
higgs  doublet  field  $\phi$   transforms  as  (1,2,2,0)   under
$G_{LR}$.

The proton decay modes allowed by the  dimension 10 operators can
be mediated  only if there  exists  $SU(3)_c$  color  non-singlet
higgs  scalars.  These  scalars  are  present  when  $G_{LR}$  is
embedded in a larger group $G_{PS} \equiv SU(4)_c \otimes SU(2)_L
\otimes  SU(2)_R$  or  other  larger  groups.  The  field  $\phi$
belongs to $(1,2,2)$ of $G_{PS}$,  but this does not give correct
mass relations \cite{pss} and one requires another field $\xi$  transforming
as  $(15,2,2)$   under   $G_{PS}$   \cite{pss,pati}.  The  fields
$\Delta_L$ and $\Delta_R$ are contained in larger representations
(10,3,1)  and  (10,1,3)  of  $G_{PS}$.  For the decay mode of the
operator  ${\cal  O}^2$  we  need  the  $SU(3)_c$  color  triplet
components of the fields $\Delta_L$ and $\xi$, which we represent
by  $\Delta_L^3$  and  $\xi^3$   respectively.  Then  the  Yukawa
couplings
\begin{equation}
{\cal  L}_{Yuk} = f_{ql} (\overline{{q_{i
\alpha L}}^c} l_{j L}) \Delta_{\alpha^* I L}^{3^*}
(\tau^I \epsilon)_{i j}
+ f_{dl} (\overline{{d_{\alpha \hat{i} R}}^c}
{l_{i d L}}^c) \xi_{\alpha^* i \hat{i} }^{3^*}
\end{equation}
(where $\hat{i}, \hat{j}, ... $ are the $SU(2)_R$ indices) and the
quartic scalar coupling
\begin{equation}
{\cal L}_{s} = \lambda \Delta_{\alpha I L}^{3}
\Delta_{\beta J L}^{3}
\xi_{\gamma i \hat{i} }^{3}
\xi_{j \hat{j} }^{1} \epsilon_{\alpha \beta \gamma}
(\tau^K \epsilon)_{I J} (\tau^K \epsilon)_{i j}
\epsilon_{\hat{i} \hat{j}}
\end{equation}
give the $(B - L)$ conserving proton decay
$P \rightarrow {e_L}^+ \nu_L {\nu_L}^c$
through the  operator  ${\cal  O}^2$ (at the  level of
standard model there is no distinction between the $\phi$ and the
$\xi^1$) through the diagram of figure 1. The amplitude for the
process is given by,
\begin{equation}
{\cal A} = \displaystyle \frac{ \lambda f_{ql}^2 f_{dl} \langle
\xi^1 \rangle }{ m_{\xi^3}^2 m_{\Delta^3}^4} .
\end{equation}
where, $\langle \xi^1 \rangle = \langle \phi \rangle = 250$ GeV.
The  mass  of   $\xi^3$   can  be  as  low  as  $\sim   100$  GeV
\cite{pss,pati}.  Then for a typical value of the quartic and the
Yukawa couplings  parameters,  $\lambda \sim 10^{-2}$ and $f \sim
10^{-3}$,  and for  $ m_{\Delta^3}  \sim 6 \times 10^{4}$ GeV we
obtain the lifetime for the proton  decay in this  particular
mode  (\ref{dkmod}) to be $4 \times 10^{31}$ years which can explain
the atmospheric neutrino anomaly.

The  mass  scale  $  m_{\Delta^3}$   need  some  explanation.  In
theories where left-right symmetry is broken at a very low energy
\footnote{The  lower bound on the  left-right  symmetry  breaking
scale comes from the LEP data to be $\sim 10^3$ GeV \cite{leppr}}
$m_R \sim 1 - 10$ TeV, one can have $m_{\Delta^3} \sim 10^4$ GeV.
Otherwise in theories  where the left-right  $D-$parity is broken
spontaneously   \cite{par},   this   scale   can   have   another
explanation.  In this  scenario the  $D$-parity  is broken by the
{\it vev} of the singlet field $\eta$ (1,1,1,0), which transforms
under  $D$ as $\eta \to - \eta$.  The  scalar  and the  fermionic
fields   transform   under   $D-$parity  as   $\Delta_{L,R}   \to
\Delta_{R,L}$  and $\psi_{L,R} \to \psi_{R,L}$,  while $\phi$ and
$\xi$  stay the same.  With  the field  $\eta$  the  lagrangian
now contains terms,
$$
{\cal L}_{\eta \Delta} = M_{\eta} \eta (\Delta_R^\dagger \Delta_R
  - \Delta_L^\dagger \Delta_L) + \lambda_{\eta}  \eta^2
  (\Delta_L^\dagger  \Delta_L + \Delta_R^\dagger \Delta_R)
$$
which can then allow a different scenario for the
masses.  The masses of the fields $\Delta_L$ and $\Delta_R$ are then
given by,
$$
m_{\Delta_L}^2 =  m_\Delta^2 - M_{\eta} \langle \eta \rangle + \lambda_{\eta}
{\langle \eta \rangle }^2 \;\; {\rm and} \;\;
m_{\Delta_R}^2 =  m_\Delta^2 + M_{\eta} \langle \eta \rangle + \lambda_{\eta}
{\langle \eta \rangle }^2 $$
where $m_\Delta$ is the mass parameter for $\Delta_{L,R}$
and generates the left-right symmetry breaking.
One can now fine tune  parameters  to get a solution
$$
\langle \eta \rangle \sim \langle \Delta_R \rangle \gg \langle
\Delta_L  \rangle\;\;\;
{\rm and}
\;\;\; M_\eta \approx m_{\Delta_R} \approx \langle \Delta_R \rangle
\approx m_{\Delta}
\;\;\;
{\rm and}
\;\;\; m_{\Delta_L} \ll  \langle \Delta_R \rangle $$
We can thus have $m_{\Delta^3} \sim m_{\Delta_L} \sim 10^4$ GeV, even
when $\langle \Delta_R \rangle$ is as large as $10^{10}$ GeV.

For  completeness  we shall now write down the  operators  in the
presence of the triplet higgs scalar  $\Delta_L$ which transforms
as (3, -2) under $SU(2)_L \times U(1)_Y$.  These are given by,

\vbox{
\begin{eqnarray}
{{\cal O}^1} & = & (\overline{{u_{\alpha a R}}^c}
u_{\beta b R}) (\overline{{d_{\gamma c R}}^c} e_{d R}) (\overline{l_{k
e L}} {l_{l f L}}^c) \epsilon_{\alpha \beta \gamma} \Delta^I_L
(\tau^I \epsilon)_{k l}
	\nonumber \\
{{\cal O}^2} & = & (\overline{{u_{\alpha a R}}^c}
u_{\beta b R})(\overline{{d_{\gamma c R}}^c}
{l_{k d L}}^c)(\overline{l_{l e L}} e_{f R}) \epsilon_{\alpha \beta
\gamma} \Delta^I_L (\tau^I \epsilon)_{k l}
	\nonumber \\
{{\cal O}^3} & = & (\overline{{u_{\alpha a R}}^c}
e_{b R}) (\overline{{d_{\beta c R}}^c} u_{\gamma d R}) (\overline{l_{k
e L}} {l_{l f L}}^c) \epsilon_{\alpha \beta \gamma} \Delta^I_L
(\tau^I \epsilon)_{k l}
	\nonumber \\
{{\cal O}^4} & = & (\overline{{u_{\alpha a R}}^c}
{l_{k b L}}^c) (\overline{{d_{\beta c R}}^c} e_{d R}) (\overline{l_{l
e L}} u_{\gamma f R}) \epsilon_{\alpha \beta \gamma} \Delta^I_L
(\tau^I \epsilon)_{k l}
	\nonumber \\
{{\cal O}^5} & = & (\overline{{u_{\alpha a R}}^c}
{l_{k b L}}^c) (\overline{{d_{\beta c R}}^c} u_{\gamma d R})
(\overline{l_{l e L}} e_{f R}) \epsilon_{\alpha \beta
\gamma} \Delta^I_L (\tau^I \epsilon)_{k l}
	\nonumber \\
{{\cal O}^6} & = & (\overline{{u_{\alpha a R}}^c}
e_{b R}) (\overline{{d_{\beta c R}}^c}
{l_{k d L}}^c)(\overline{l_{l e L}} u_{\gamma f R}) \epsilon_{\alpha \beta
\gamma} \Delta^I_L (\tau^I \epsilon)_{k l}
	\nonumber \\
{{\cal O}^7} & = & (\overline{{u_{\alpha a R}}^c}
e_{b R}) (\overline{{q_{i \beta c L}}^c} q_{j \gamma d L}) (\overline{l_{k
e L}} {l_{l f L}}^c) \epsilon_{\alpha \beta \gamma} \epsilon_{i j}
\Delta^I_L (\tau^I \epsilon)_{k l}
	\nonumber \\
{{\cal O}^8} & = & (\overline{{u_{\alpha a R}}^c}
{l_{k b L}}^c) (\overline{{q_{i \beta c L}}^c} q_{j \gamma d L})
(\overline{l_{l e L}} e_{f R}) \epsilon_{\alpha \beta \gamma}
\epsilon_{i j} \Delta^I_L (\tau^I \epsilon)_{k l}
	\nonumber \\
{{\cal O}^9} & = & (\overline{{q_{i \alpha a L}}^c}
l_{j b L}) (\overline{{d_{\beta c R}}^c} u_{\gamma d R}) (\overline{l_{k
e L}} {l_{l f L}}^c) \epsilon_{\alpha \beta \gamma} \epsilon_{i j}
\Delta^I_L (\tau^I \epsilon)_{k l}
	\nonumber \\
{{\cal O}^{10}} & = & (\overline{{q_{i \alpha a L}}^c}
l_{j b L}) (\overline{{d_{\beta c R}}^c} l_{k d L}}^c) (\overline{l_{l
e L}} {u_{\gamma f R}) \epsilon_{\alpha \beta \gamma} \epsilon_{i j}
\Delta^I_L (\tau^I \epsilon)_{k l} \label{op3}
\end{eqnarray} }

\noindent These  operators can give many possible  diagrams which
will  allow  three  lepton  decay  mode of the  proton  which can
explain  the   atmospheric   neutrino   problem.  The  first  two
operators ${\cal O}^{1,2}$ has a antisymmetric combination of two
$u'$s, which gives charmed meson decay modes for the proton.  All
the  remaining  operators  can, in  principle,  give  rise to the
desired decay modes of the proton (\ref{dkmod}).

To  summarize,  we have given an operator  analysis for the three
lepton decay mode of the proton.  The proton decay process  which
can explain the  atmospheric  neutrino  problem is not allowed by
dimension 9 operators.  This is allowed when a $SU(2)_L$  doublet
or a triplet field acquire a {\it vev}, which requires  dimension
10 operators  and  sometimes  can be  comparable  to  dimension 9
operators.  We write down a set of possible  operators  which can
give  this  proton  decay  mode.  In   particular   a  left-right
symmetric  model which can allow this  proton  decay mode with an
amplitude  of  right  magnitude  for  a  natural  choice  of  the
parameters.

{\bf  Acknowledgment}  We would like to thank the NSERC of Canada
for an International  Scientific  Exchange Award.  One of us (US)
would  like to thank  Prof  J.C.  Pati  for  updating  him on the
recent interest on this problem and many helpful discussions.  US
also thanks Profs S.  Pakvasa and T.  Yanagida for discussions on
the problem.

\vskip 1in
\noindent {\bf Figure Caption}

\vskip .3in
\noindent {\bf Figure 1} \hskip .3in Diagram
giving $P \rightarrow {e_L}^+ \nu_L {\nu_L}^c$.

\newpage
\middlespace

\end{document}